\documentclass[10pt,prc,aps,psfig,twocolumn]{revtex4}
\usepackage{graphicx}
\tighten
\begin{document}
\draft
\title{The impact of charge symmetry and charge independence breaking on the properties
of neutrons and protons in isospin-asymmetric nuclear matter 
}
\author{F. Sammarruca, Larz White, and B. Chen}
\address{Physics Department, University of Idaho, Moscow, ID 83844, U.S.A}
\date{\today}

\email[F. Sammarruca: ]{fsammarr@uidaho.edu}

\begin{abstract}
We investigate the effects of charge independence and charge symmetry breaking in neutron-rich 
matter. We consider neutron and proton properties in isospin-asymmetric
matter at normal densities as well as the high-density neutron matter equation of state and the
bulk properties of neutron stars. 
We find charge symmetry and charge independence breaking effects to be very small.
\end{abstract}
\maketitle

\section{Introduction}
Charge independence (CI) of the nuclear force signifies that, in a state of total isospin equal to one (the 
only accessible to two neutrons or two protons), 
the neutron-proton ($np$), neutron-neutron ($nn$), and proton-proton ($pp$) interactions are equal, after electromagnetic
effects have been removed.
Charge independence breaking (CIB) refers to slight violations of this property. On the other hand,
charge symmetry breaking (CSB) refers only to differences between the $nn$ and the $pp$ interactions.
Clearly, CSB also breaks CI, while the opposite is not necessarily true.           
Basically, 
charge independence is a statement of invariance under rotations in isospin space, whereas CS is a special case
of CI and amounts to invariance under a rotation from $T_z=1$ to $T_z=-1$. 
A thorough review of charge dependence can be found in Ref.~\cite{MNS}. 

At a fundamental level, charge dependence of the nuclear force is understood in terms of differences
between the masses of up and down quarks and electromagnetic interactions among the quarks. 
At the hadronic level, CSB manifests itself through mass differences between hadrons of the same 
isospin multiplet, meson mixing, and other mechanisms such as meson-photon exchanges.

Both CSB and CIB are seen most clearly in the $^1S_0$ scattering lengths, which have the values
\cite{MNS,How,GT,anp1,anp2,anp3} 
\begin{displaymath}
a_{nn} = -18.9 \pm 0.4 \;\mbox{fm} \; ,       
a_{pp} = -17.3 \pm 0.4 \; \mbox{fm} \; ,     
\end{displaymath}
\begin{equation}
a_{np} = -23.74 \pm 0.02 \; \mbox{fm} \; .   
\end{equation}
These values for the scattering lengths imply that the $nn$ interaction is slightly more attractive
than the $pp$, and the (T=1) $np$ interaction is more attractive than both $pp$ and $nn$. 
The difference between $a_{nn}$ and $a_{pp}$ is a measure of CSB, whereas 
\begin{equation}
\Delta = \frac{1}{2}(a_{nn} + a_{pp}) - a_{np}                
\end{equation}
is typically taken as a measure of CIB. 
Thus, the singlet scattering length of nucleon-nucleon (NN) scattering shows small CSB (about 10\%) and larger CIB.
It has been shown that both nucleon mass splitting \cite{LM} and $\rho^0-\omega$ mixing \cite{CB}             
can fully explain charge symmetry breaking in the singlet scattering length. Thus, different observables
need to be identified which might reveal sensitivity to the mechanism used to describe CSB. 

In this paper, CSB/CIB are introduced in two different ways. In one case, we use the high-precision, charge-dependent NN potential known as 
the ``CD-Bonn" potential \cite{CDB}. The latter is based upon the Bonn full model \cite{Bonnfull} and 
reproduces the predictions of the comprehensive model for CSB and CIB in all partial waves with 
$J\le 4$. The other model uses $\rho - \omega$ mixing for CSB, with the mixing Lagrangian as in 
Ref.~\cite{MM01}. 

Within our recent efforts to investigate diverse aspects of the nuclear equation of state (EoS) \cite{FS10}, 
our main purpose in this paper is to explore the effects of charge dependence and asymmetry of the nuclear force 
in a highly neutron-rich environment, such as isospin-asymmetric nuclear matter (IANM), 
and how the effect evolves as a function of density and/or isospin asymmetry.      

Isospin-asymmetric nuclear matter simulates the interior of a heavy nucleus with unequal densities of protons and neutrons.
The equation of state of (cold) IANM is then a function of density as well as the relative concentrations 
of protons and neutrons. 
The recent and fast-growing interest in IANM stems from its close connection to the physics of neutron-rich nuclei, or,
more generally, asymmetric nuclei, including the very ``exotic" ones known as ``halo" nuclei. 
The equation of state of IANM is also the crucial input for the structure equations of
compact stars, and thus establishes the connection between nuclear physics and compact astrophysical systems. 
Hence, in-depth studies of IANM are important as well as timely.

In symmetric nuclear matter, CSB would cause a splitting of the single-particle potential, 
$\Delta U=U_p-U_n$. Some                                                                                   
calculations within the mean field approximation \cite{Kim+} find that CS
is gradually restored in nuclear matter in $\beta$-equilibrium, whereas                           
CSB becomes strongly enhanced away from $\beta$-equilibrium because of the large   
nucleon isovector density. 
Medium effects on CSB in neutron matter 
were investigated in neutron matter \cite{Ishi} and found to reduce
the mass of a neutron star (composed of neutrons only) by as much as 35\% as compared to the one 
resulting from a standard relativistic calculation with no CSB effects.

In the next Section, we will describe how CSB and CIB effects are 
included in the IANM environment. We will then present and discuss some results, with a particular eye
on selected isovector properties, specifically the single-proton and single-neutron potentials, the 
neutron-proton mass splitting, and the symmetry potential. Neutron star masses and radii               
will also be shown.                                                             
Conclusions and future plans are summarized in the last Section. 

\section{Description of the calculation }                                                
\subsection{The CSB and CIB mechanisms through mass splitting }                                                
CD-Bonn is a high-precision charge-dependent one-boson-exchange potential whose charge dependence
is based upon the predictions of the Bonn full model \cite{Bonnfull}.
The charge symmetry breaking due to nucleon mass splitting was investigated in Ref.~\cite{LM} 
with the Bonn full model.                  
Considerable CSB was found to be generated by $2\pi$ and $\pi-\rho$ diagrams. In addition to the 
well-known CSB in the singlet scattering length, non-negligible CSB effects were observed in higher
partial waves, especially $P$ and $D$ waves. Apparently, those are crucial for a quantitative account of
the Nolen-Shiffer \cite{NS} anomaly seen in the energies of neighboring mirror nuclei \cite{MPM}.

Concerning CIB, pion mass splitting was found to be a major cause for it, in the one-pion as well
as multi-pion exchanges. In fact, the contribution from multi-pion exchanges to CIB is about 50\% 
of the one originating from one-pion exchange \cite{EM,LM2}.   

CD-Bonn consists of three different potentials,$V_{pp}$,
$V_{nn}$, and $V_{pn}$, 
for the $pp$, $nn$, and $pn$ cases, respectively. 
First, $V_{pp}$ is constructed through fits to the $pp$ data. Next, starting from 
$V_{pp}$ and replacing the mass of the proton with the mass of the neutron,                     
 the $nn$ singlet scattering length is reproduced. The                
$V_{np}$ in the T=1 channel is obtained using the proper charge-dependent one-pion exchange plus
a slight adjustment of the $\sigma NN$ coupling to account for the remaining CIB. 
The predictions by CD-Bonn are 
\begin{displaymath}
a_{nn} = -18.9680 \;   \mbox{fm} \; ,  \\
a_{pp} = -17.4602 \; \mbox{fm} \; ,\\
\end{displaymath}
\begin{equation} 
a_{np} = -23.7380 \; \mbox{fm} \; .   
\end{equation}

\subsection{Application to isospin-asymmetric nuclear matter }                                                

The framework we use to calculate the EoS of IANM \cite{FS10} naturally lends itself to the application of potentials 
which are different for different nucleon pairs. 

Here we use the Brueckner-Hartree-Fock (BHF) description of nuclear matter.
The scattering equation for two nucleons 
in nuclear matter reads, schematically 
\begin{eqnarray}
G_{ij} = V_{ij} + V_{ij}Q_{ij} G_{ij} \; , 
\end{eqnarray}                    
where $G_{ij}$ is the in-medium reaction matrix 
($ij$=$nn$, $pp$, or $np$).                                              
$Q_{ij}$ is the Pauli operator, which prevents scattering to occupied $nn$, $pp$, or $np$ states.            
In the case of CD-Bonn, the scattering equation (in free space) is derived from the Blankenbecler and Sugar
\cite{BbS} reduction of the of the four-dimensional Bethe-Salpeter equation \cite{BS}, and is formally
identical to the non-relativistic 
Lippman-Schwinger equation.

The goal is to determine self-consistently the nuclear matter single-particle potential   
which, in IANM, will be different for neutrons and protons. 
We have, for neutrons,
\begin{equation}
U_n = U_{np} + U_{nn} \; , 
\label{un}
\end{equation}
and for protons
\begin{equation}
U_p = U_{pn} + U_{pp} \, , 
\label{up}
\end{equation}
where each of the four pieces on the right-hand-side of Eqs.~(\ref{un}-\ref{up}) signifies an integral of the  appropriate $G$-matrix ($nn$, $pp$, or $np/pn$). The splitting of the single-particle potential manifest in  
Eqs.~(5-6) is due to the presence of two different Fermi momenta for neutrons and protons, regardless 
charge effects. In addition, 
charge symmetry and charge independence breaking can then be incorporated in a natural way using the appropriate potential in each of the 
integrals above, whereas in standard charge-independent calculations the same potential (namely, the $np$ potential) is used for all nucleon pairs. 
Thus, for an arbitrary level of isospin asymmetry, the effects we include are both CIB and CSB effects.
On the other hand, in nearly pure neutron matter, where the proton Fermi momentum approaches zero,
$U_{np}$ in Eq.~(5) and $U_{pp}$ in Eq.~(6) will approach zero. The $U_{pn}$ term in Eq.~(6) becomes
the (very attractive) potential of a proton impurity, but it has essentially zero weight in the 
final averaging of neutrons and protons due to the vanishing proton Fermi momentum. Therefore, 
using the CD-Bonn $nn$ potential in the second term on the right-hand side of Eq.~(5) instead of the $np$ potential (as usually done) is to say that the mechanism
under consideration is primarily CIB.

\section{Results and discussion}                                                
\subsection{Some neutron and proton properties in IANM}                                                

All charge-dependent effects shown in this subsection are included through the use of the  
appropriate version of the CD-Bonn potential in Eqs.~(5-6). 

We begin with the splitting of the neutron and proton effective masses in IANM, see Fig.~1, 
as a function of the neutron excess parameter, $\alpha =\frac{\rho _n - \rho _p}{\rho_n+\rho_p}$, in terms of neutron and proton densities. The large splitting between the effective
masses, which vary almost linearly and in opposite directions, is of course due to the presence of (increasing) isospin asymmetry in the medium. The 
much smaller effect, noticeable only on the right side mostly through the differences between the green 
and the blue lines, is due to charge dependence.             
Comparison between the predictions on the left side and on the right side of the figure suggests
that these effects increase slightly with total density, and that they are larger for neutrons.

Figures 2 and 3 show charge-dependent effects on the single-neutron potential as a function of the momentum and fixed density. 
Comparison between the predictions on the left side and on the right side of both figures indicates that 
charge-dependence effects 
increase only very little as the neutron excess parameter changes from 0.5 to nearly 1. 
Taking CIB properly into account makes $U_n$ more repulsive because the $nn$ interaction is less attractive than
the $np$ one.

\begin{figure}
\centering            
\vspace*{-1.0cm}
\hspace*{-2.0cm}
\scalebox{0.3}{\includegraphics{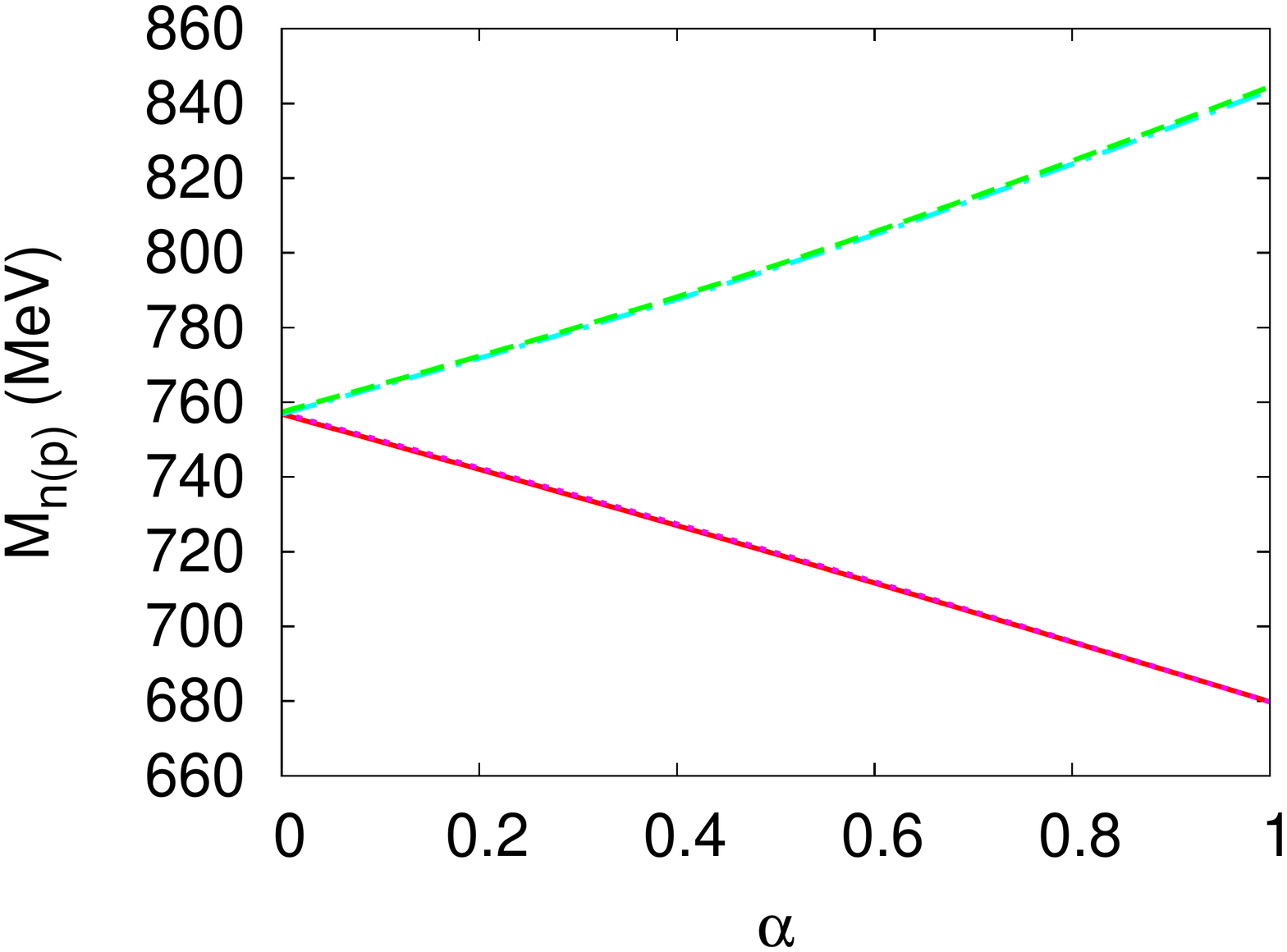}}
\scalebox{0.3}{\includegraphics{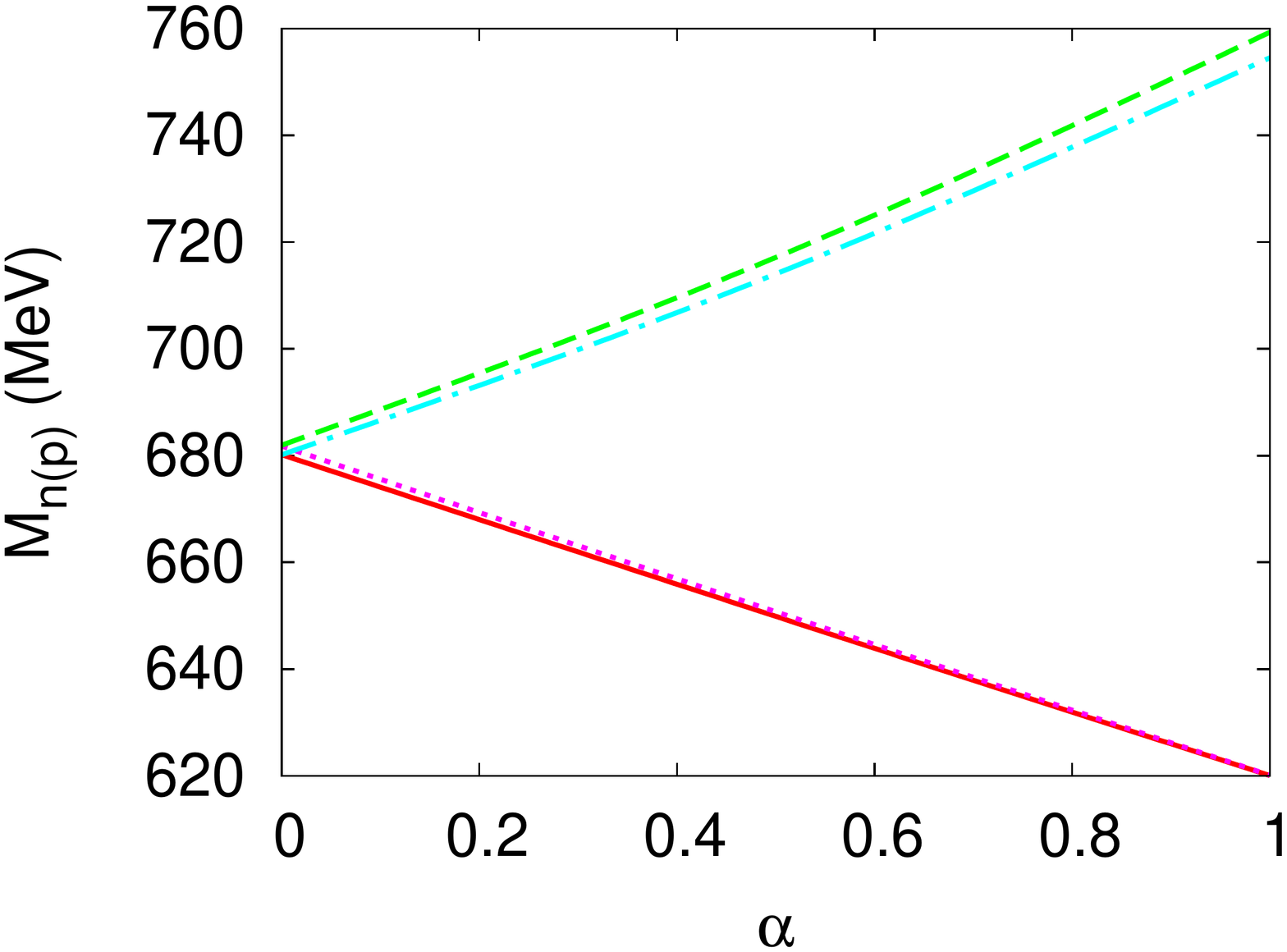}}
\vspace*{0.001cm}
\caption{Neutron (green and blue lines) and proton (red and pink lines) effective masses in IANM as a function of the neutron excess
parameter and for fixed density. The Fermi momentum is equal to 1.4 fm$^{-1}$ and 
2 fm$^{-1}$ in the upper and lower frame, respectively. 
The green dashed and the pink dotted lines show the predictions with CSB/CIB.                                             
} 
\label{one}
\end{figure}

\begin{figure}
\centering            
\vspace*{-1.0cm}
\hspace*{-2.0cm}
\scalebox{0.3}{\includegraphics{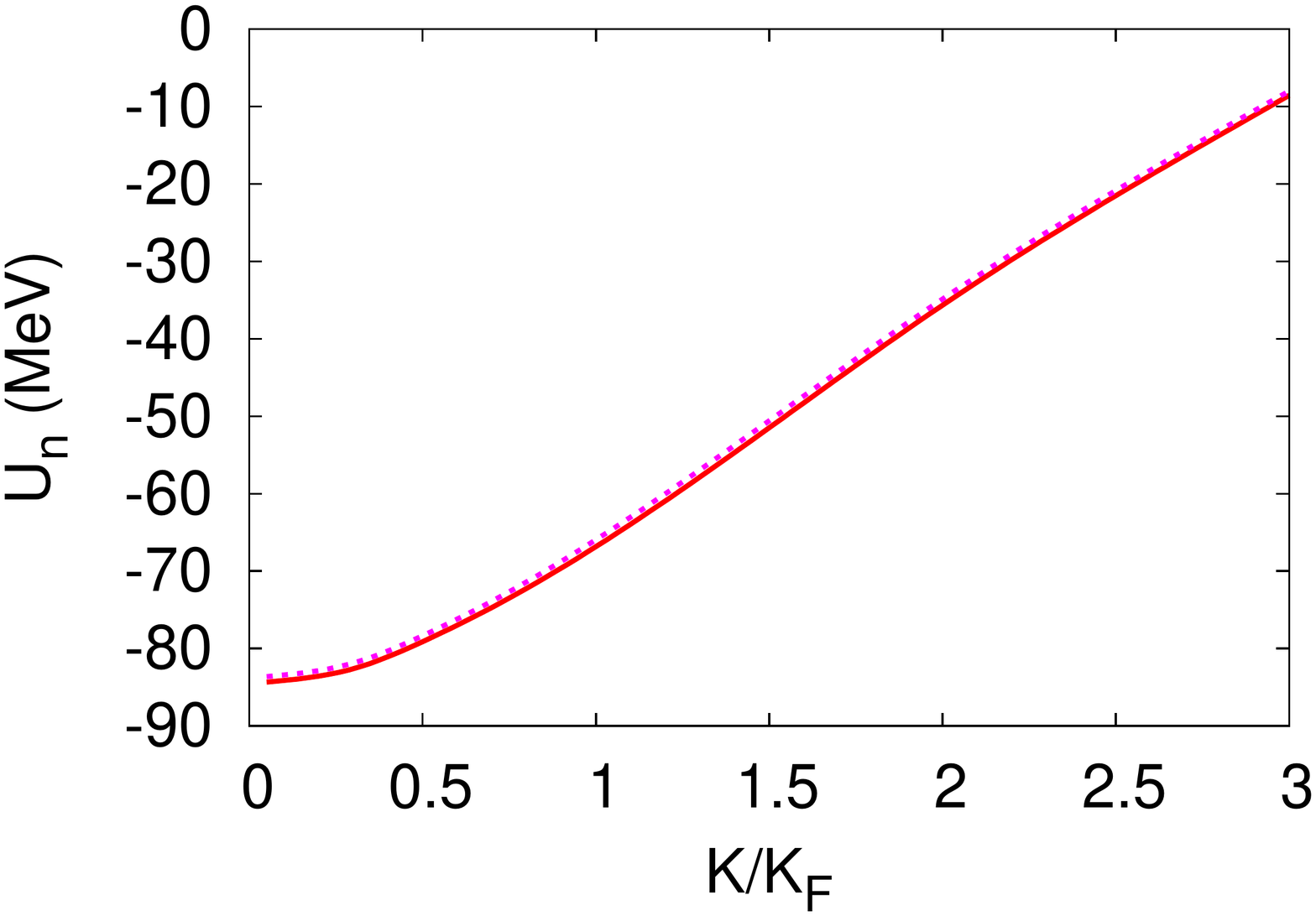}}
\scalebox{0.3}{\includegraphics{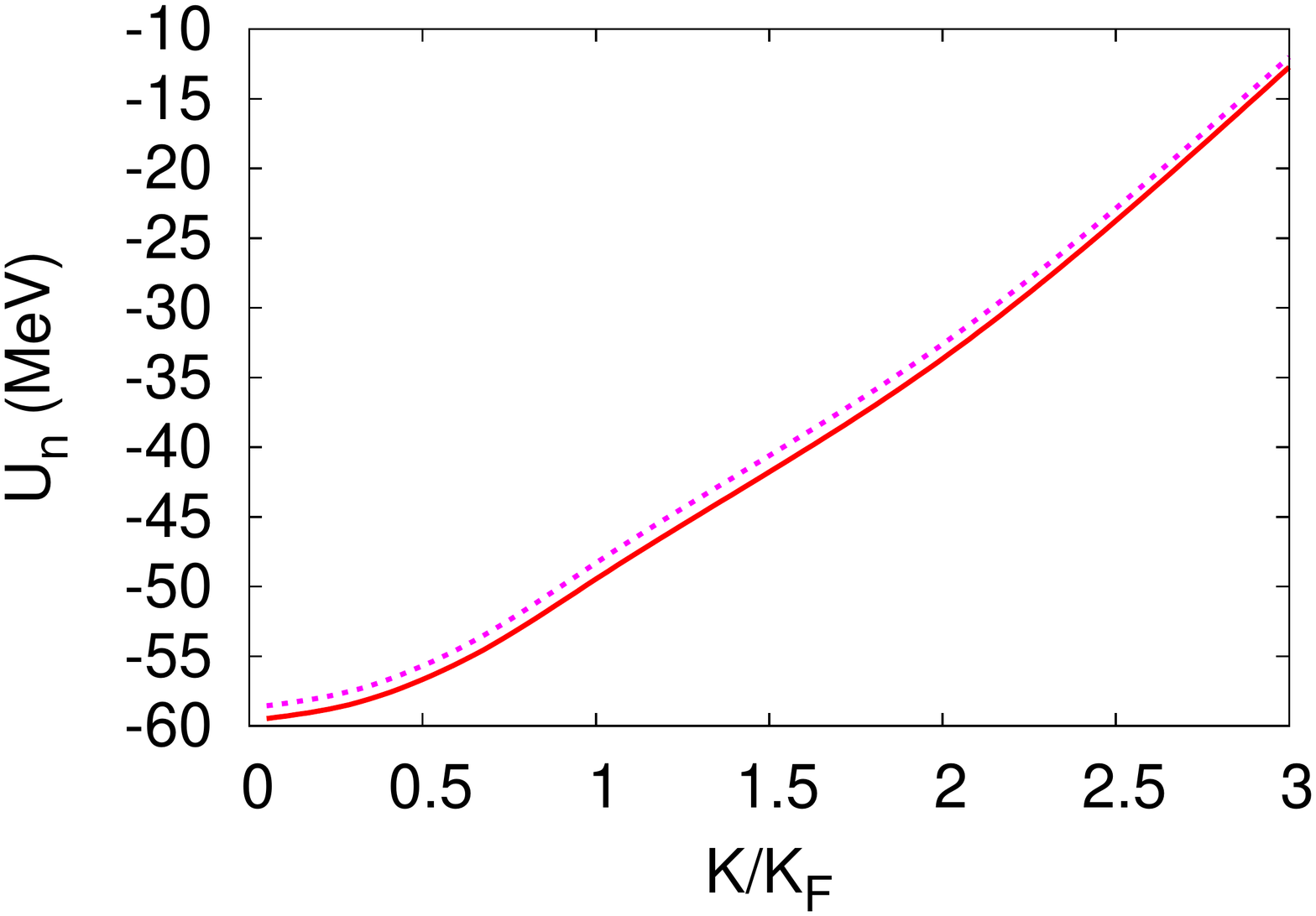}} 
\vspace*{0.001cm}
\caption{The single-neutron potential in IANM as a function of momentum (in units of the 
Fermi momentum) for fixed density and neutron excess parameter. 
In both frames the Fermi momentum is equal to 1.4 fm$^{-1}$, whereas the 
neutron excess parameter is equal to 0.5 in the upper frame and nearly 1 (almost all neutrons) in the lower frame. 
The pink dotted line, almost indistinguishable from the solid red in the left panel, corresponds to the 
calculation including charge-dependent effects. 
} 
\label{two}
\end{figure}

\begin{figure}
\centering            
\vspace*{-1.0cm}
\hspace*{-2.0cm}
\scalebox{0.3}{\includegraphics{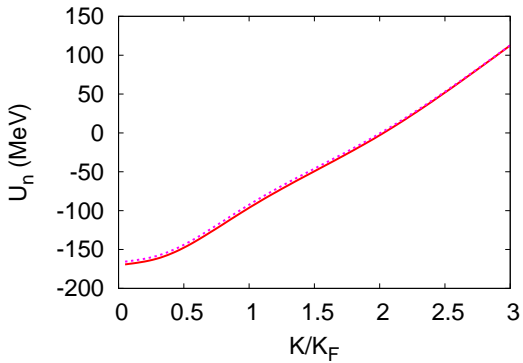}}
\scalebox{0.3}{\includegraphics{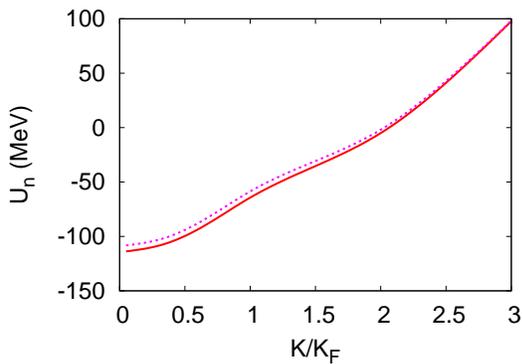}} 
\vspace*{0.3cm}
\caption{As in Fig.~2, but with a                                                              
Fermi momentum equal to 2 fm$^{-1}$.           
} 
\label{three}
\end{figure}

\begin{figure}
\centering            
\vspace*{-1.0cm}
\hspace*{-2.0cm}
\scalebox{0.3}{\includegraphics{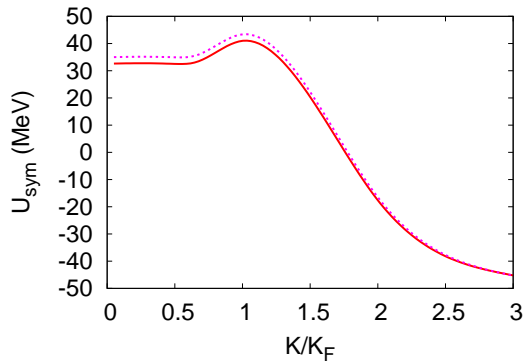}}
\scalebox{0.3}{\includegraphics{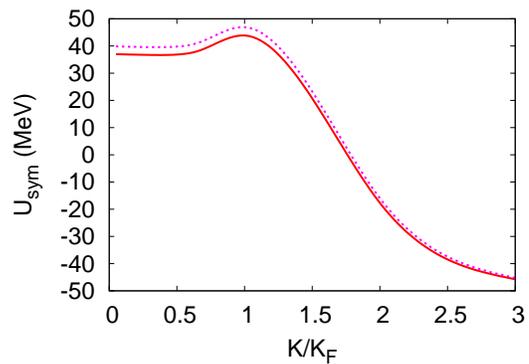}} 
\vspace*{0.01cm}
\caption{The symmetry potential without (solid red) and with (dotted pink) CSB/CIB effects as a function 
of momentum and fixed Fermi momentum, equal to 
2 fm$^{-1}$. The     
neutron excess parameter is equal to 0.5 in the upper frame and nearly 1 in the lower one. 
} 
\label{four}
\end{figure}
The symmetry potential, defined as 
\begin{equation}
U_{sym} = \frac{U_n - U_p}{2 \alpha} \; , 
\end{equation}
plays a crucial role in separating the dynamics of neutrons and protons in collisions of 
asymmetric systems. It can be related to the isovector component of the nuclear optical potential,
and thus is a valuable quantity when seeking constraints to the EoS of IANM.
In Fig.~4, we show the symmetry potential as a function of momentum and for two 
different values of the isospin asymmetry. Because $U_n$ and $U_p$ are approximately linear with 
respect to $\alpha$, the symmetry potential should be nearly independent of $\alpha$, which is 
confirmed in Fig.~4. 
Again, the dotted line displays predictions obtained with charge-dependent effects. 
We observe that, although small, the difference is noticeable, particularly at the lowest momenta. 

\subsection{Masses and radii of neutron stars }                                                
From the results shown in the previous Section, one might conclude that CSB and CIB effects in nuclear matter, 
even if very neutron rich, are quite small. Nevertheless, it is insightful to take the (charge-dependent)
EoS to the very high densities probed by compact stars and calculate neutron star bulk properties. 
We will show these predictions as obtained with 
both the mass splitting and the $\rho - \omega$ mixing models.

In the latter, 
the charge asymmetric nuclear force is evaluated by adding a ``mixed" 
 $\rho-\omega$ exchange.                                       
 The mixing matrix element was extracted in Ref.~\cite{CB} from the $e^+ e^- \rightarrow \pi ^+ \pi ^-$
annihilation process and found to be
\begin{equation}
<\rho ^0|H_m|\omega> = -0.00452 \pm 0.0006 \; \;  \mbox{GeV}^2 \; . 
\end{equation}
Within the 
 $\rho-\omega$ mixing model, the CSB potential is developed starting with the $pp$ CD-Bonn potential,
which is the most accurate due to the large base of high-quality $pp$ data, 
and then constructing the $nn$ potential by adding to the $pp$ potential a term that contains the difference        
$V_{\rho \omega} = V_{\rho \omega}(nn) - 
V_{\rho \omega}(pp)$ generated by 
 $\rho-\omega$ mixing.                                                                                  
The reader is referred to Ref.~\cite{MM01} for more details.

All predictions shown in this Section are calculated with:
1) No consideration of charge dependence or charge asymmetry (curve labeled    
``No CD" in Figs.~5-8); 2) CIB and CSB effects as included in CD-Bonn (the curve labeled
``CD, mass splitting" in Figs.~5-8); and 3) $\rho-\omega$ mixing for the CSB sector. 
At this point, it is useful to
recall the comments made following Eqs.~(5-6) concerning which kind of charge effects we may 
potentially detect in a systems of only neutrons.

In Figs.~5-6 we see CIB effects on the energy per particle and the pressure in neutron matter. The effect of CIB is small
but noticeable, particularly at the higher densities, However, the predictions from the two models   
are essentially indistinguishable. 
The inclusion of CIB renders the EoS slightly more repulsive. This is reasonable, as the interaction 
between two identical nucleons                   
described in terms of a charge-independent NN potential (typically $np$), is less repulsive
than one designed specifically for two neutrons.

The impact of CIB on the mass-radius relation of a neutron star (comprised of only neutrons)
is shown in Fig.~7. We note in passing that,        
 in spite of the attractive nature of CD-Bonn, and, particularly, the use of the conventional 
BHF calculation of the EoS, one might have expected even smaller maximum masses than those shown in  
Fig.~7. However,                                        
we must keep in mind that only neutrons are involved here (i.e. the T=1 potential). The over-attraction expected from
CD-Bonn comes mainly from 
the T=0 potential.     

Back to CIB considerations, 
 the maximum mass changes from 1.58 to 1.6 solar masses due to CIB, a variation of only 1.3\%. 
In Fig.~8, the mass of the star is shown as a function of the central energy density for the three 
models, showing that the central density corresponding to the maximum mass is approximately 
the same in all cases.

In closing this Section, we take note of the findings from 
Ref.~\cite{Ishi}, where a large                                                                         
reduction of the star mass, as much as 35\% around normal density, was reported as a consequence of CSB. In       
that work, $\rho-\omega$ mixing takes place through a baryon loop, as initially proposed by Piekarewicz ans Williams \cite{PW}, 
and is subjected to medium effects. 
In contrast, we        
attribute $\rho-\omega$ mixing to electromagnetic coupling, see Eq.~(8).

\begin{figure}
\centering            
\vspace*{-1.1cm}
\hspace*{-2.0cm}
\scalebox{0.35}{\includegraphics{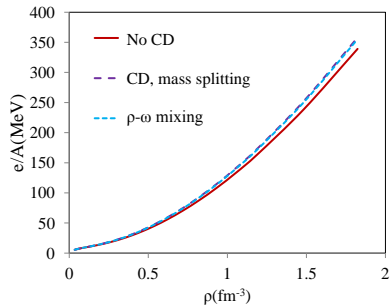}}
\vspace*{-4.0cm} 
\caption{The energy per particle in neutron matter for the three models 
discussed in the text.} 
\label{five}
\end{figure}

\begin{figure}
\centering            
\vspace*{1.102cm}
\hspace*{-2.0cm}
\scalebox{0.35}{\includegraphics{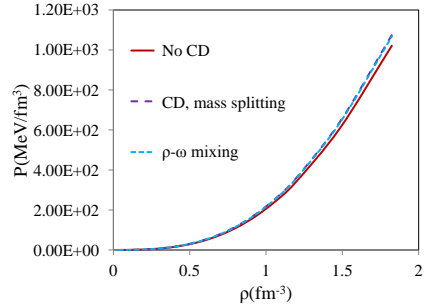}} 
\vspace*{-4.0cm}
\caption{The pressure in neutron matter for the three models 
discussed in the text.} 
\label{six}
\end{figure}

\section{Summary and conclusions}                                                           

Symmetries and the mechanisms responsible for their breaking are issues of fundamental importance
as they help our understanding of the nuclear force at its very core. 

In this paper, 
we have investigated the impact of CIB and CSB of the NN interaction on the splitting between neutron and proton properties
in IANM. We find such impact to be very small, even at very high neutron relative concentrations.
Given the chief role of the symmetry potential in collisions of isospin-asymmetric systems, it may be useful
to keep in mind that there is some small but noticeable sensitivity at low momentum. 

We then moved to high-density neutron matter and calculated neutron star masses and radii. 
Consistently with the EoS becoming slightly more repulsive with considerations of CIB, the 
star mass is found to increase by a few percent. 

We noted that, from calculations where (density-dependent) nucleon-antinucleon loops allow the transition between isoscalar
and isovector mesons, a strong enhancement of CSB is predicted in neutron matter.
This is definitely an important issue to be further explored, as a large CSB effect
in stellar matter would impact the cooling mechanism of the star. 

In the near future, we plan to explore if and how the presence of 
relativistic effects impact CIB and CSB in stellar matter. 
To that end, we will      
produce a charge-dependent relativistic potential (such as a ``charge-dependent Bonn B")
suitable for applications 
within the relativistic Dirac-Brueckner-Hartree-Fock approach.

\begin{figure}
\centering            
\vspace*{1.0cm}
\hspace*{-2.0cm}
\scalebox{0.4}{\includegraphics{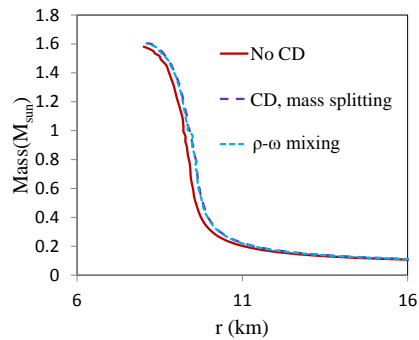}}
\vspace*{-5.0cm}
\caption{The mass of a neutron star {\it vs.} the radius for the three models under consideration.}     
\label{seven}
\end{figure}

\begin{figure}
\centering            
\vspace*{-0.5cm}
\hspace*{-2.0cm}
\scalebox{0.4}{\includegraphics{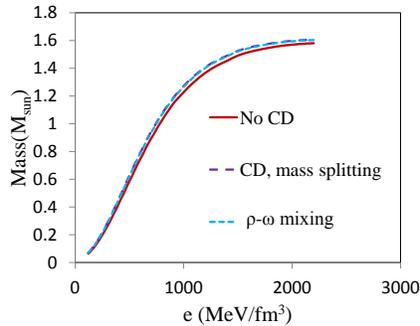}}
\vspace*{-5.0cm}
\caption{The mass of a neutron star {\it vs.} the central density for the three models under consideration.}     
\label{eight}
\end{figure}
\begin{center}
{\bf Acknowledgments}                                                           
\end{center}
\vspace{-10.0cm}
Support from the U.S. Department of Energy under Grant No. DE-FG02-03ER41270 is 
acknowledged.                                                                          
We are grateful to F. Weber for providing his computer code for the calculation of neutron star
proprties.

\end{document}